\documentclass[aip,apl,reprint,showpacs,showkeys,superscriptaddress]{revtex4-1}

\usepackage{graphicx}
\usepackage{amsmath}

\begin{document}

\title{Spin transition in Gd$_3$N@C$_{80}$, detected by low-temperature on-chip SQUID technique}

\author{L. Chen}
\affiliation{Department of Physics and The National High Magnetic Field Laboratory, Florida State University, Tallahassee, Florida 32310, USA}

\author{E. E. Carpenter}
\affiliation{Department of Chemistry, Virginia Commonwealth University, 1001 W. Main Street, Richmond, VA 23284, USA}

\author{C.S. Hellberg}
\affiliation{Code 6390, Center for Computational Materials Science, Naval Research Laboratory, Washington, DC 20375, USA}

\author{H. C. Dorn}
\affiliation{Department of Chemistry, Virginia Tech, Blacksburg, VA 24061, USA}

\author{M. Shultz}
\affiliation{Department of Chemistry, Virginia Commonwealth University, 1001 W. Main Street, Richmond, VA 23284, USA}

\author{W. Wernsdorfer}
\affiliation{Institut N\'{e}el, associ\'{e} \`{a} l'UJF, CNRS, BP 166, 38042 Grenoble Cedex 9, France}

\author{I. Chiorescu}
\affiliation{Department of Physics and The National High Magnetic Field Laboratory, Florida State University, Tallahassee, Florida 32310, USA}

\date{\today}

\begin{abstract}
We present a magnetic study of the Gd$_3$N@C$_{80}$ molecule, consisting of a Gd-trimer via a Nitrogen atom, encapsulated in a C$_{80}$ cage. This molecular system can be an efficient contrast agent for Magnetic Resonance Imaging (MRI) applications. We used a low-temperature technique able to detect small magnetic signals by placing the sample in the vicinity of an on-chip SQUID. The technique implemented at NHMFL has the particularity to operate in high magnetic fields of up to 7 T. The Gd$_3$N@C$_{80}$ shows a paramagnetic behavior and we find a spin transition of the Gd$_3$N structure at 1.2~K. We perform quantum mechanical simulations, which indicate that one of the Gd ions changes from a $^8S_{7/2}$ state ($L=0, S=7/2$) to a $^7F_{6}$ state ($L=S=3, J=6$), likely due to a charge transfer between the C$_{80}$ cage and the ion.
\end{abstract}


\keywords{SQUID magnetometery, on-chip devices, Gd$_3$N@C$_{80}$ molecule, spin transition}

\maketitle

\begin{figure}
\includegraphics[width=\columnwidth]{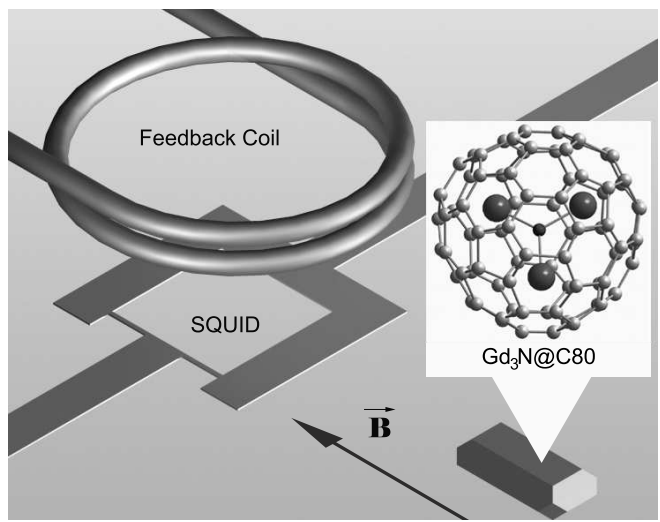}
\caption{The sketch of the on-chip SQUID setup. A feedback coil is placed above an on-chip SQUID and the magnetic field is oriented parallel to it. The Gd$_3$N@C$_{80}$ sample (not at scale) is placed near the on-chip SQUID loop. The structure of the Gd$_3$N@C$_{80}$ molecule is presented in the insert.}
\label{fig1}
\end{figure}

The metal-containing endohedral fullerenes molecules have drawn attention as they exhibit unusual material properties associated with charge transfer between the metal cores and the carbon cage.\cite{stevenson_nature1999} The encapsulated metal cores are protected from the solvent or other agents by the carbon cage, which leads to potential biological applications. Specifically, the trigadolinium fullerene Gd$_3$N@C$_{80}$ (in Fig~\ref{fig1}) has attracted wide interest, because it is identified as a safe candidate as contrast-enhancing agent in Magnetic Resonance Imaging; also, it can be implanted with other ligands for medical applications\cite{shu_biochem2009}. The carbon cage separates the interior metal atoms from the environment, preventing their leakage and accumulation into the human body. Also, vibrational modes~\cite{Burke_PRB2010} indicate a bond formation between the cage and the Gd$_3$N. Studies showed~\cite{qian_jap2007,qian_prb2007} that the Gd$_3$N cluster is bound to the C$_{80}$ cage by a large binding energy of 13.63~eV which ensures that Gd atoms are not likely to break out of the cage. The three Gd atoms are bonded through their delocalized \emph{s} and \emph{d} states, with the \emph{s} and \emph{p} states of C atoms. This leaves the spin magnetic moment of the \emph{f} electrons almost intact, with no appreciable spin density on the cage.  This compound can improve the MRI relaxation enhancement by a factor of ~20, leading to a new class of MRI tumor delineation techniques.\cite{qian_prb2007}  Such developments towards novel medical applications, require a fundamental understanding of the electronic and magnetic properties of these endohedral metalofullerenes. In this paper, we present a study of the spin properties of the Gd$_3$N core cluster by on-chip Superconducting Quantum Interference Device (SQUID) measurement technique.

Magnetic flux detectors using SQUID devices feature very high sensitivity and operate at low temperatures.~\cite{clarke_squid} Unlike commercial SQUIDs, where the sample is coupled to the detector via a pickup coil, the on-chip SQUID method couples the sample directly, by placing it on the same chip as the SQUID, for increased sensitivity~\cite{wernsdorfer_sst}. The loop of a SQUID is interrupted by two Josephson junctions (see Fig.~\ref{fig1}), which can be of planar-tunneling type~\cite{clarke_squid} or nano-bridges~\cite{anderson_prl1964} which are constrictions of a thin niobium film and are much thinner than tunneling junctions. Thin films have a significantly larger in-plane critical field, compared to the bulk value~\cite{tinkhanm_superconductivity}. Consequently, the nano-bridge junctions are less affected by a large magnetic field applied along the film plane. By aligning the magnetic field with high precision in the plane of a Nb SQUID, the on-chip SQUID technique employed here extends the observation of quantum magnetic phenomena into a higher magnetic field range. Therefore, this improved technique is very promising in gathering more information about effects such as spin state transitions and entanglements, the anisotropy-induced tunneling gap\cite{wernsdorfer_europhylett2000}, phonon bottleneck effects ~\cite{chen_apl2006,chen_erophylett2009} and others. 

SQUID operation require a precise alignment of the magnetic field $\vec{B}$. We can rotate $\vec{B}$ in SQUID's plane by combining fields generated by a 3-axis magnet $\vec{B}=\vec B_x+\vec B_y+\vec B_z$. Initially, the SQUID plane is roughly aligned along $B_z$. We can tune the direction of the field along the crossing between the SQUID plane and the $xz$ plane by applying a small amount of $B_x$, which has been set roughly normal to the SQUID plane (similar approach applies for the $yz$ plane, if using the $B_y$ component). The on-chip SQUID is made of a Nb film of only 5.5~nm thickness ensuring a superconducting state even for 7~T of in-plane magnetic field~\cite{chen_nanotech2010}. The SQUID loop has a 2$\times$2 $\mu m^2$ area and contains two parallel nano-bridge Josephson junctions. This technique allows us to place a small amount of Gd$_3$N@C$_{80}$ particles in the vicinity of the SQUID loop for an increased magnetic coupling between sample and SQUID. The entire device is thermally anchored to the mixing chamber of a dilution refrigerator to ensure an efficient cooling. 

\begin{figure}
\includegraphics[width=\columnwidth]{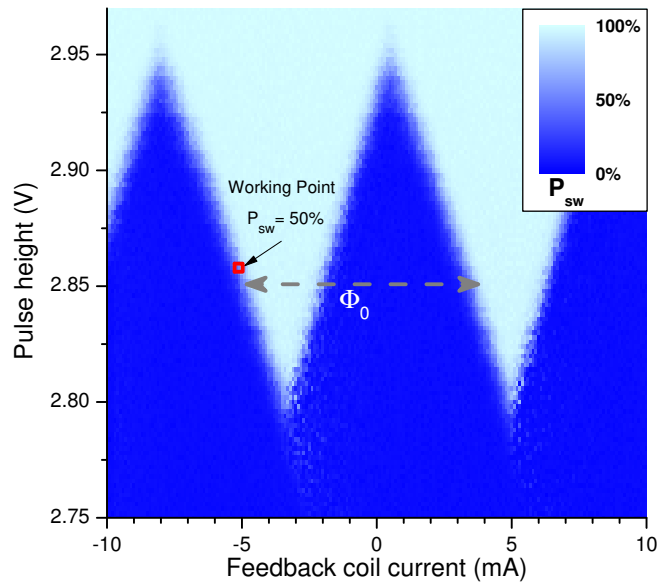}
\caption{Contour plot showing $P_{sw}$ as a function of the feedback current and the height of the switching pulse applied to SQUID. The sharp transition from low to high switching probability are a characteristic of SQUID devices. The detector is kept at a working point (here $P_{sw}=50\%$) by the feedback mechanism, relaying information on the compensated external flux, coming from a nearby sample. }
\label{fig2}
\end{figure}

The detection is done by analyzing the output voltage following a series of current pulses, sent to the SQUID. If the current pulse is sufficiently strong, the SQUID switches into the normal state, generating detectable voltage pulses. Statistically, the ratio between the number of output voltage pulses and that of the applied current pulses, gives $P_{sw}$, the switching probability~\cite{chiorescu_science2003}. SQUID switching into the normal state generates heat, which we minimize by operating the SQUID in pulse mode. In our experiment, current pulses are generated by a pulsed voltage source with an in-series resistor. The switching probability $P_{sw}$ of the SQUID increases with the current amplitude flowing through the junctions. This can be observed in the contour plot of Fig.~\ref{fig2}, showing sharp transitions between low and high switching probabilities. We define the switching current $I_{sw}$ as the current corresponding to $P_{sw}=50\%$ and Fig.~\ref{fig2} shows a typical periodic modulation as a function of the magnetic flux in the SQUID loop, with a period of $\Phi_0=h/2e$ ($h$ is Planck's constant and $e$ the electron charge). 

Real time (``on-the-fly") magnetic measurements are performed using a feedback mechanism, described below. An "working point" for the SQUID is chosen, for instance at the middle point in the modulation curve (see Fig.~\ref{fig2}). During the measurement, a current pulse with a height corresponding to the working point is sent to the SQUID. It is important to note that the variation of the probability near the working point is very sharp, which is the key point behind the high flux sensitivity of a SQUID. We use a small superconducting coil placed above the SQUID (see Fig~\ref{fig1}), to generate a flux compensating any flux changes caused by the magnetic sample.  The feedback coil current is proportionally related to the change in the magnetic moment of the sample. This method locks the SQUID at its working point and performs a feedback measurement.

\begin{figure}
\includegraphics[width=\columnwidth]{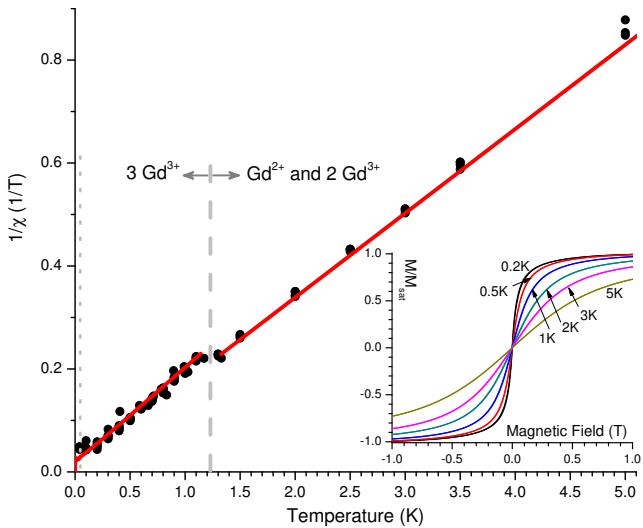}
\caption{The inverse of the initial susceptibility (black dots) as a function of temperature, calculated from normalized magnetization curves (see insert). The continuous line is a fit based on Eqs.~\ref{exchangeJ} and \ref{susc}. The insert shows such curves, taken at different temperatures. The different slopes before and after the transition temperature (dashed line at $\approx$1.2~K) indicate a spin transition, due to a charge transfer onto the Gd$_3$ system. The dotted line at very low temperatures ($<0.1$~K) indicates a region where the $J=1/2$ ground state of the Gd trimer, starts to become dominant (see text). Insert: Reversible, normalized magnetization curves measured at different temperatures, as indicated in the figure.}
\label{fig3}
\end{figure}

The described setup is used to measure the magnetization of Gd$_3$N@C$_{80}$ molecules, conglomerated in a small particle, while sweeping the applied field from negative to positive saturation ($-1$ to $+1$~T). The magnetic field was set at a sufficiently low sweeping rate to ensure a thermal equilibrium magnetization process. The high symmetry of the C$_{80}$ cage does not favor the existence of magnetic anisotropy in the Gd$_3$N system, as confirmed by our reversible magnetization curves (see below). The collective spin momentum of Gd$_3$N core cluster exhibit a paramagnetic behavior, as expected also from theoretical calculations.~\cite{qian_prb2007}  In the insert of Fig.~\ref{fig3} magnetization curves at different temperatures are presented. No hysteresis is observed for all temperatures, confirming that the sample is completely paramagnetic. The magnetization curves are flattened with the increase of temperature, which gives a smaller initial susceptibility $\chi$. To obtain the initial susceptibility, we normalize the magnetization curves to the saturation value (at very low temperature and 1~T) and then take the derivative at zero field. A plot $1/\chi$ vs. $T$ is shown in Fig.~\ref{fig3}. We observe a linear dependence but with an abrupt change in slope at $\sim 1.2$~K.

We fit the susceptibility data using a Heisenberg model for the Gd moments.
The external magnetic field $\vec{B}$ couples to the total moment $\vec{J}_{\rm i}$ of Gd ion $i$, but the exchange interaction is mediated through the spin operators $\vec{S}_{\rm i}$. Using $\vec{S}_{\rm i} = (g_{\rm i}-1) \vec{J}_{\rm i}$, with $g_{\rm i}$ the Land\'{e} g-factor of ion ${\rm i}$, we can write the entire Hamiltonian as:
\begin{equation}
H =
 - \sum_i g_{\rm i}       \mu_{\rm B} \vec{J}_{\rm i}  \cdot  \vec{B}
+
\sum_{i<j} J_{{\rm ij}} (g_{\rm i}-1)(g_{\rm j}-1)\vec{J}_{\rm i} \cdot  \vec{J}_{\rm j}, 
\label{exchangeJ} 
\end{equation} 
where $\mu_{\rm B}$ is the Bohr magneton, $J_{\rm ij}$ is the exchange coupling between spins $i$ and $j$ ($i,j=1..3$). We compute the susceptibility simply by enumerating all possible states using the quantum numbers $\vec{J'}= \vec{J}_1 + \vec{J}_2$ and $\vec{J} =\vec{J'} +\vec{J}_3$. The thermal averaging of all eigenstates, with energies $E_{\rm k}$ and
magnetizations
$M_{\rm k}=\partial E_{\rm k}/\partial B$,
gives the susceptibility as:
\begin{equation}
\chi=\frac{\beta}{Z}\sum_{\rm k}{M^2_{\rm k}e^{-\beta E_{\rm k}}}, 
\label{susc}
\end{equation}
where $\beta=1/k_{\rm B}T$, $k_{\rm B}$ is Boltzmann's constant and $Z=\sum_{\rm k}e^{-\beta E_{\rm k}}$ is the partition function.

In both the experiment and the fit, the susceptibility is normalized to the saturation magnetization at low temperatures (see the slope for $T<$1.2~K in Fig.~\ref{fig3}).  For the fit, the saturation magnetization from the three Gd ions is given by $\sum_{i=1...3} \mu_{\rm B}g_{\rm i} \sqrt{J_{\rm i}(J_{\rm i}+1)}$.

We find that the low-temperature susceptibility is well described by three Gd$^{3+}$ ions in $^8S_{7/2}$ states ($L=0, S=7/2$) interacting via equal exchange couplings $J_{\rm ij} = J\approx$ 7.6~mK. For T $>1.2$~K, it appears that one Gd changes to a 2+ state with a $^7F_{6}$ configuration ($L=S=3, J=6$). This shows that one electron is transferred from the C$_{80}$ cage onto one of the Gd ions, leading to an overall spin transition for the Gd$_3$ system. The model also agrees with a flattening of the $1/\chi$ vs. $T$ curve at very low temperatures (below $\sim 0.1-0.2$~K) when the exchange couplings $J_{\rm ij}$ becomes noticeable, and the $J=1/2$ ground state of the system begins to dominate the susceptibility. 

In conclusion, we studied the magnetization dynamics of Gd$_3$N@C$_{80}$, a molecular system useful as a high contrast agent for MRI, as a function of temperature. The magnetic studies are performed by using a sensitive on-chip SQUID technique. The initial susceptibility data are fitted with a quantum mechanical model, using all the states of the Gd trimer. At low temperatures, the data are explained by 3 Gd$^{3+}$ ions in state $^8S_{7/2}$ weakly coupled antiferromagnetically. A spin transition of the Gd$_3$ is observed at 1.2~K  explained by a charge transfer of one electron from the C$_{80}$ cage, onto one of the Gd ions. The later becomes a Gd$^{2+}$ ion, in a state $^7F_{6}$. 
 
We acknowledge support from the NSF Cooperative Agreement Grant No. DMR-0654118, the NSF-CAREER grant No. DMR-0645408 and the Alfred P. Sloan Foundation. WW thanks the financial support from the ANR-PNANO projects MolNanoSpin ANR-08-NANO-002 and the ERC Advanced Grant MolNanoSpin 226558.

\end{document}